# MULTI-STANDARD PROGRAMMABLE BASEBAND MODULATOR FOR NEXT GENERATION WIRELESS COMMUNICATION

Indranil Hatai[1] and Indrajit Chakrabarti[2]

[1]Department of Electronics and Electrical Communication Engineering, Indian Institute of Technology, West Bengal, India
indranilh@cse.iitkgp.ernet.in

[2]Department of Electronics and Electrical Communication Engineering, Indian Institute of Technology, West Bengal, India
indrajit@ece.iitkgp.ernet.in

## ABSTRACT

*Considerable research has taken place in recent times in the area of parameterization of software defined radio (SDR) architecture. Parameterization decreases the size of the software to be downloaded and also limits the hardware reconfiguration time. The present paper is based on the design and development of a programmable baseband modulator that perform the QPSK modulation schemes and as well as its other three commonly used variants to satisfy the requirement of several established 2G and 3G wireless communication standards. The proposed design has been shown to be capable of operating at a maximum data rate of 77 Mbps on Xilinx Virtex 2-Pro University field programmable gate array (FPGA) board. The pulse shaping root raised cosine (RRC) filter has been implemented using distributed arithmetic (DA) technique in the present work in order to reduce the computational complexity, and to achieve appropriate power reduction and enhanced throughput. The designed multiplier-less programmable 32-tap FIR-based RRC filter has been found to withstand a peak inter-symbol interference (ISI) distortion of -41 dBs.*

## 1. INTRODUCTION

The standards of wireless communication system are changing very rapidly and the new standards have been growing up along with existing old standards. The current mobile phones basically support only a fixed number of standards. This emerging technology has evolved towards the concept of SDR [1-8]. The SDR concept first appeared in the military area [9], and was later implemented in the civil communication area [10]. SDR uses programmable digital devices to perform the signal processing necessary to transmit and receive the baseband information at intermediate frequency (IF). A mobile handset is required to support implementation of physical layer protocols of several cellular communication standards in order to fulfil the requirements of "communication anywhere anytime within a single terminal". Devices such as digital signal processors (DSP's) and field programmable gate array (FPGA's) are endowed with signal processing functionality required for hardware realization of these communication protocols. SDR enables the consumers as well as manufacturers to upgrade their product to satisfy different communication protocol, simply by the up-gradation of the software, which will configure the hardware at any time according to the needs. The reconfiguration of the hardware in SDR can be done when the product is being shipped or when a new standard emerges or even when a call is in progress. It results in a shorter development time and cheaper production cost due to higher volumes. All of these benefits provide opportunities for SDR in wireless communication market.

Although DSP's as well as FPGA's may be used to implement an SDR baseband protocol, a DSP needs more time to reconfigure the hardware in comparison with the FPGA [11-12]. The present paper proposes an FPGA-based programmable baseband modulator (PBM) which is capable of performing the QPSK modulation scheme as well as its other three commonly used variants, at a maximum data rate of about 77 Mbps. The paper is organized as follows, Section 2 is about the previous research on the baseband processing of SDR system, In Section 3; the basic modulation scheme has been identified for different wireless communication standards for realizing a common baseband modulator which is useful for 2G and 3G air interface. The architecture of the PBM and its component has been discussed in Section 4. Section 5 provides the FPGA implementation results. Section 6 draws the conclusions.

## 2. PREVIOUS RESEARCH

A prevalent trend witnessed lately in wireless communication systems has been to support all the existing multiple radio access protocols along with an eye to improving the inter-system compatibility. This has made way for a product which can include multiple standards into a single device. This is possible by implementing the concept of re-configurability in the radio system architecture. In modern-day handheld wireless devices, the baseband processing part is commonly performed by the inflexible ASICs, which are superior in performance. But according to the need, one can support different standards/protocols at different times in a single device only by reconfiguring the programmable devices (FPGA, DSP). So, several methods have evolved to implement reconfigurable radio or software defined radio (SDR) architecture. F Jondral [13] first introduced the idea of parameterization or dynamic reconfiguration in the digital radio system architecture. He has demonstrated how the idea of parameterization can help produce an SDR system which can integrate 2G as well as 3G standard in a common platform. Wiesler et al [14] also shows the roadmap for the SDR by investigating the parameter list which is being used at the time of software radio reconfiguration to implement all the important mobile standards. In between, several silicon solutions for implementing the SDR by the best possible hardware choices have been find out in [15]. In this paper the authors have suggested FPGA+DSP solution for the generalized SDR scheme, which is suitable for several mobile air interfaces. The low-size, low-power SDR prototype system developed for multi-mode and multi-task software radio by Harada et al [16] can support several standards e.g. global positioning system (GPS) navigation system, vehicle information and communication system (VICS), electronic toll collection system (ETC), AM/FM radio broadcasting service, FM multiplex broadcasting system and several modulation schemes such as BPSK, QPSK, GMSK, ASK, and $\pi/4$ QPSK by downloading the software. Moreover, the multi-task algorithm allows a user to obtain multiple services at a time. Later, a real-time SDR test-bed for baseband processing of the physical layer of wireless standards implementation on a general purpose processor (GPP) has been described in [17]. Several techniques have been explored for the implementation of common baseband processing. A total cost approach has been adopted for baseband processing in SDR architecture in [18]. In this literature the authors have evaluated the quantitative cost factor for the baseband processing implementation in different reconfigurable architecture. In [19], M. Imran Anwar et al have proposed a new design framework for common baseband processing where after exploring the algorithmic and architectural design specifications of 3G and 4G systems, they have been integrated in an innovative extensible hardware. A parametric design effort has been made by T. Becker et al [20] for reconfigurable SDR architecture. In this paper, it has been demonstrated by a case study of reconfigurable FIR filter how the degree of parallelism affects the reconfiguration time and performance in run-time reconfigurable FPGAs. U. Ramacher [21] has pointed out two alternative approaches by which the existing as well as the evolving cell phone standards keeps changing the functions of a system architect in response to the market requirements. L. Mingfu et al [22] have presented software radio based universal digital baseband modulator architecture for baseband processing part. In this context, novel approach has been taken to develop a low-area, low-power, high-speed common baseband processor for the next generation wireless communication device.

## 3. DIFFERENT MODULATION SCHEMES

The proposed baseband modulator can be integrated into the baseband processing structure of the wireless communication standard listed in Table 1. The concept of SDR is based on the ability of a single terminal to support multiple wireless communication standards on the same platform. F. Jondral [23] introduced the concept of the parameter controlled software radio structure by combining different modulators and demodulators under the same transceiver architecture. The PBM proposed in the present paper can perform any one of the four modulation schemes, namely (A) Quadrature Phase Shift Keying (QPSK), (B) Differentially Encoded QPSK (DQPSK), (C) DQPSK with $\pi/4$ phase shift, (D) Orthogonal QPSK (OQPSK).

Table 1. Prescribed data rate and modulation schemes of some wireless standards

| Standard | Data Rate | Modulation Type |
| --- | --- | --- |
| PHS/PACS | 384 Kbps | $\pi/4$ DQPSK |
| IS-54/IS-136 | 30 Kbps | $\pi/4$ DQPSK |
| IS-95 | 1.2288 Mbps | QPSK/OQPSK |
| IMT-2000/UMTS | 3.6864 Mbps | QPSK |
| GPS | 2 Mbps | QPSK |
| PDC | 1.6 Mbps | $\pi/4$ DQPSK |
| DVB-S | 9.14 Mbps | QPSK |
| DAB-T | - | DQPSK |
| SDARS | - | QPSK(Satellite) |
| Zigbee | 20, 40, 250 Kbps | OQPSK |

### 2.1. QPSK (Quaternary Phase Shift Keying)

Several communication standards use the QPSK modulation as a main baseband modulation scheme for its robustness against noise and very high data rate. In QPSK, two data channels modulate the carrier which results an increase in bandwidth efficiency. Transition of the original message signal causes a change in the phase of the carrier. The phase of the carrier changes to equally spaced value, namely $0, \pi/2, \pi$ and $3\pi/2$, where each value corresponds to a unique pair of message bits. QPSK produces a constant envelope signal which is given as

$$S_{QPSK}(t) = \sqrt{2P} e^{j(2\pi f_c t + \theta(t) + \theta_c)} \quad (1)$$

where $P$ is the transmitted power, $f_c$ is the carrier frequency in Hz, $\theta_c$ is the carrier phase and $\theta(t)$ is the data phase that takes on equally spaced value

$$\theta(t) = (i-1)\pi/2 \quad \text{where} \quad i = 1,2,3,4$$

### 2.2. Differentially Encoded QPSK

DQPSK modulation scheme is widely used in the various communication standards because of its power and bandwidth efficiency. In DQPSK systems, the input binary sequence is first differentially encoded and then modulated using QPSK modulator. The differentially encoded symbol is derived based on the present and on the last encoded symbol. The Boolean expression of the nth differentially encoded in-phase (I) and quadrature (Q) bits are represented as,

$$I'_n = I_n \overline{I}_{n-1} \overline{Q}'_{n-1} + Q_n \overline{I}'_{n-1} Q_{n-1} + \overline{I}_n I_{n-1} Q_{n-1} + \overline{Q}_n I_{n-1} \overline{Q}'_{n-1} \qquad (2)$$

$$Q'_n = Q_n \overline{I}'_{n-1} \overline{Q}'_{n-1} + \overline{I}_n \overline{I}'_{n-1} Q_{n-1} + \overline{Q}_n I_{n-1} Q_{n-1} + I_n I_{n-1} \overline{Q}'_{n-1} \qquad (3)$$

### 2.3. DQPSK with $\pi/4$ Phase Shift

The $\pi/4$ DQPSK modulation scheme is widely used because of its superior spectral efficiency. It is a special case of QPSK which offers a compromise between OQPSK and QPSK in terms to allow maximum phase change of 1350compared to 1800 for QPSK and 900 for OQPSK. $\pi/4$ DQPSK signals are differentially encoded to facilitate easier implementation of differential detection or coherent detection with phase ambiguity in the recovered carrier. The complex envelope of a $\pi/4$ DQPSK modulated signal with the symbol duration T is

$$S_{DQPSK}(t) = \sum_{n=0}^{\infty} \exp[\theta(n)] \qquad (4)$$

where the phase of the complex symbols $Z_n = \exp[\theta(n)]$ are

$$\theta(n) = \theta(n-1) + \Delta\theta(n) \qquad (5)$$

### 2.4. Orthogonal QPSK

High power amplifiers (HPA) near to or at saturation are widely used in the digital radio transmission to gain the maximum power efficiency. The $180^0$ phase change in QPSK modulation scheme forces the envelope to pass through a zero crossing. Due to this when the band-limited QPSK signal passes through the HPA results in an increase of ISI by means of spectral spreading  OQPSK modulation scheme is similar to the QPSK modulation scheme except the quadrature data stream is delayed with respect to the inphase data stream by one bit period. OQPSK modulation scheme eliminates the $180^0$ phase transition by making the even and odd data stream orthogonal to each other. It has been adopted in the IS-95 standard due to its superior ISI reduction ability, though maintaining the same bandwidth occupancy as compared to the simple QPSK.

### 4. PROPOSED PROGRAMMABLE BASEBAND MODULATOR

The proposed architecture focuses on the hardware multiplexing technique, where the hardware can be efficiently shared and utilized between different operations. This in turn leads to reduced area and power consumption at a reasonable data rate. By integrating several of modulation schemes in one common structure, and reusing the hardware for different standards one can minimize the amount of program memory in the processor, can save valuable development time, and reduce the development costs. However in a reprogrammable multi-standard baseband processor, hardware multiplexing can often achieve comparable power consumption and lower silicon area than a fixed function circuit [24-25]. The architecture of the proposed programmable multi-standard baseband processor using hardware multiplexing technique is as depicted in Figure 1.

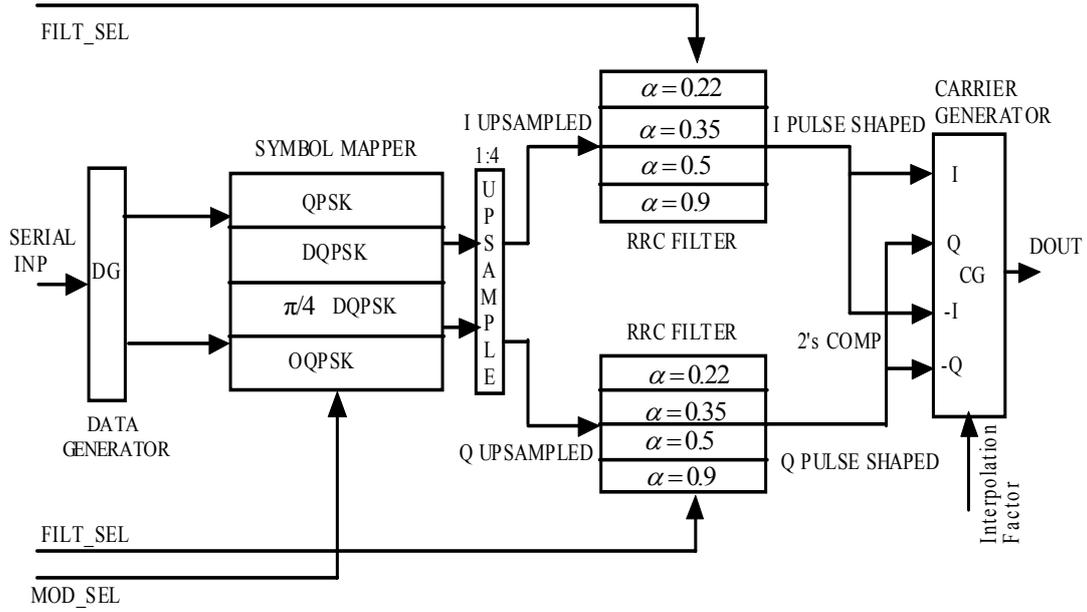

Figure 1. Block Diagram of Proposed Programmable Baseband Modulator

The key component of the multi-standard baseband processor, as shown in Figure 5 includes the (A) Serial-Parallel converter (shown as DG), (B) Symbol Mapper, (C) Upsampler, (D) Root Raised Cosine Filter as a pulse shaping filter, and (E) Carrier Generator (shown as CG) block.

### 4.1. Data Generator (DG)

A dummy data acquisition module accepts the digitized data from the Analog to Digital Converter (ADC) and then passes through a serial to parallel converter for dividing the incoming message bit stream into the even and odd bit stream. To accommodate the OQPSK scheme a delay of one bit period has been placed in the quadrature data stream path. After that the generated even and odd bit streams passes through the symbol mapper block.

### 4.2. Symbol Mapper

Symbol Mapper block contains the four different IQ mapper blocks for four different modulation schemes for generating the carrier signal. It takes the even and the odd bit stream from the data generator block and generates a stream of mapped I and Q bit streams each of 3-bit length. The symbol mapper block consists of four blocks described below which can be chosen by the MOD_SEL parameter.

### 4.2.1. QPSK Mapper

This block accepts the even and odd bit stream and generates a phase of the current bit sequence according to the Table 2 which causes the carrier to shift by $90^0$ or $180^0$ or $270^0$. The shift in the carrier generates the symbols belonging to the set {-1, 0, 1} for both I and Q data streams. The block diagram of the QPSK mapper block is as shown in Figure 3.

Table 2. Phase shift information of QPSK

| Information Bits | | Phase Shifts (Radians) |
|---|---|---|
| *I* | *Q* | |
| 0 | 0 | 0 |
| 0 | 1 | $\pi/2$ |
| 1 | 0 | $\pi$ |
| 1 | 1 | $3\pi/2$ |

### 4.2.2. DQPSK Mapper

The block schematic of the differential encoder is shown in Figure 2. The differentially encoded symbol is derived from the present and one of the last encoded symbols. Suppose the $n^{th}$ symbol, $(n-1)^{th}$ encoded symbol and the $n^{th}$ encoded symbol be represented as $\{I_n\ Q_n\}$, $\{I'_{n-1} Q'_{n-1}\}$ and $\{I'_n Q'_n\}$ respectively then the encoding scheme for the DQPSK modulation is carried out according to (2) and (3).

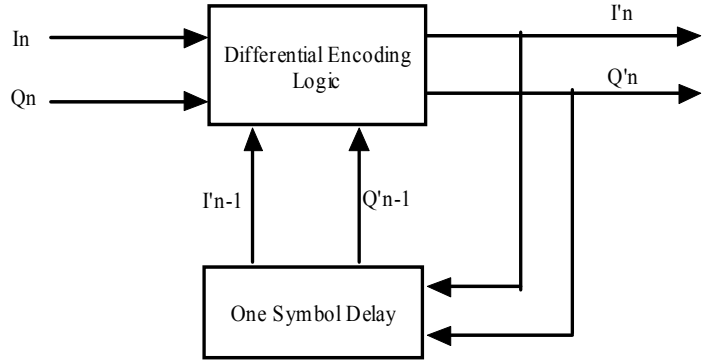

Figure 2. Block Schematic of Differential Encoding Scheme of DPQSK Modulator

### 4.2.3. $\pi/4$ DQPSK Mapper

It takes two bit at a time and assigned to one of the four possible differential phases as shown in Table 3. The actual information is placed into the differential phase of two successive symbols. The symbols corresponding to the possible state space of {-1, -0.707, 0, 0.707, 1} will be generated for both of I and Q signal set. The block diagram of the π/4DQPSK mapper block is as shown in Figure 3.

Table 3. Phase shift information of $\pi/4$ DQPSK

| Information Bits | | Phase Shifts (Radians) |
|---|---|---|
| *I* | *Q* | |
| 0 | 0 | π/4 |
| 0 | 1 | 3π/4 |
| 1 | 0 | 5π/4 |
| 1 | 1 | 7π/4 |

### 4.2.4. OQPSK Mapper

In OQPSK mapper the even and the odd signals are offset in their alignment by one bit period (half symbol period). For, due to the transition in the input symbol at any time only one of the two bit streams can change the value. Therefore the maximum phase difference of $90^0$ can occur with the possible state space of {-0.707, 0.707} for I and Q component. The block diagram of the OQPSK mapper block is as shown in Figure 3.

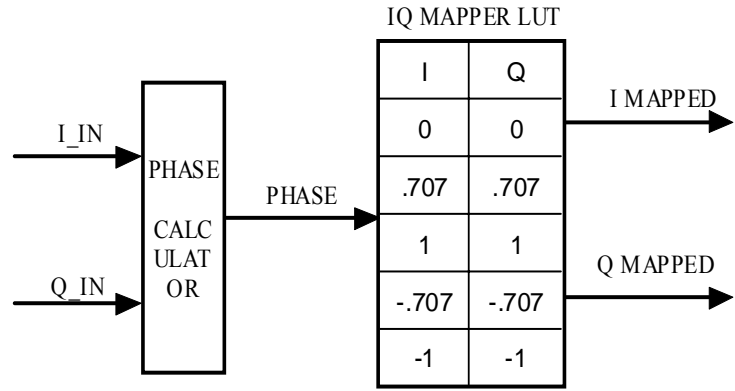

Figure 3. Block Diagram of IQ Mapper for QPSK, $\pi/4$ DQPSK, OQPSK

### 4.3. Upsampler

Upsampler block takes the mapped data stream input from the previous block and will generate the upsampled I and Q data by a factor of 4. This upsampler block consists of a chain of registers connected serially which acts as a shift register. The seven prior symbols are loaded into the shift register and generate the addresses for the distributed arithmetic based RRC filter as shown in Figure 5(a).

### 4.4. Pulse Shaping Filter

Inter-symbol interference is the main problem in any digital modulation technique which leads to increase in the probability of the receiver making an error in detecting a symbol. In mobile communication, spectral shaping at the IF stage using RRC pulse shaping filter [26-27] can be used to reduce the inter-symbol effects as well as the spectral width of the modulated signal. Real-time implementation of RRC filter needs a wide eye-opening to fight against the probability of high bit error rate (BER) by disallowing the timing jitter at the sampling instant. The transfer function of a raised cosine filter [28] is given by

$$H(t) = \frac{\sin\left(\frac{\pi t}{T_s}\right)}{\pi t} \bullet \frac{\cos\left(\frac{\pi \alpha t}{T_s}\right)}{1 - \left(\frac{4\alpha t}{2T_s}\right)^2} \tag{6}$$

where $\alpha$ is the roll-off factor which ranges between 0 and 1. The magnitude response of the RRC filter with roll-off factor 0.35 is as shown in Figure 4.

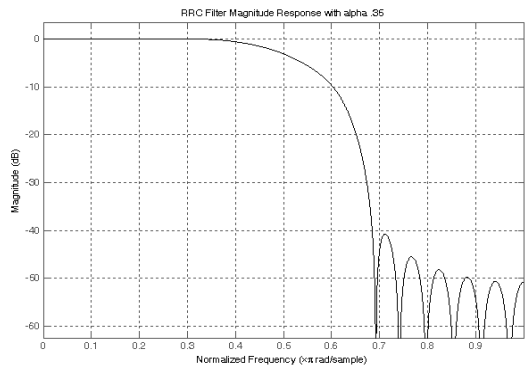

Figure 4. RRC Filter Magnitude Response with $\alpha = 0.35$

Figure 4 justifies that the RRC filter with roll-off factor 0.35 can withstand the peak ISI distortion of -41dB. Depending on the different applications several researches has been carried over on the various architecture of RRC filter implementation [29-32]. The increasing demand of the high speed with low computational complexity pulse shaping filter, in the signal processing field has become the key point of the research in RRC filter implementation. Typical implementation of RRC FIR filter is based on multiply and accumulate approach where the computational complexity, is a function of the multipliers and adders [33-34]. To reduce the computational complexity another multiplier less architecture based on DA technique has been proposed by White [35], where the only the number of adders connected serially determines the computational complexity. In the proposed design implementation of the DA based parameter controlled RRC-FIR filter for low computational complexity, power optimization and enhanced throughput has been done.

RRC filter block contains the root raised cosine filter with the different value of roll-off factor or excess-bandwidth factor ($\alpha$). The value of $\alpha$ used here are 0.22, 0.35, 0.5, 0.9 according to the need for different standards e.g. IS-95, UMTS, WCDMA. In the present work a 32-tap RRC finite impulse response (FIR) filter with different values of $\alpha$ has been designed in MATLAB for generating the coefficients. For the hardware implementation of the RRC filter, each coefficient is multiplied with 4096 for the integer representation. The filter also functions as 1:4 interpolators which mean it will produce 20-megasamples/sec while taking the input data at 5-megasamples/sec. The RRC filter has been implemented using DA technique where the multiplication is achieved by pre-storing the sum of partial products for all the terms within an equation and addressing the look-up table (LUT) with the bits of all the input variables. The upsampler block generates the address for the corresponding LUT. The block diagram of the RRC filter using the DA technique is as shown in Figure 5(b).

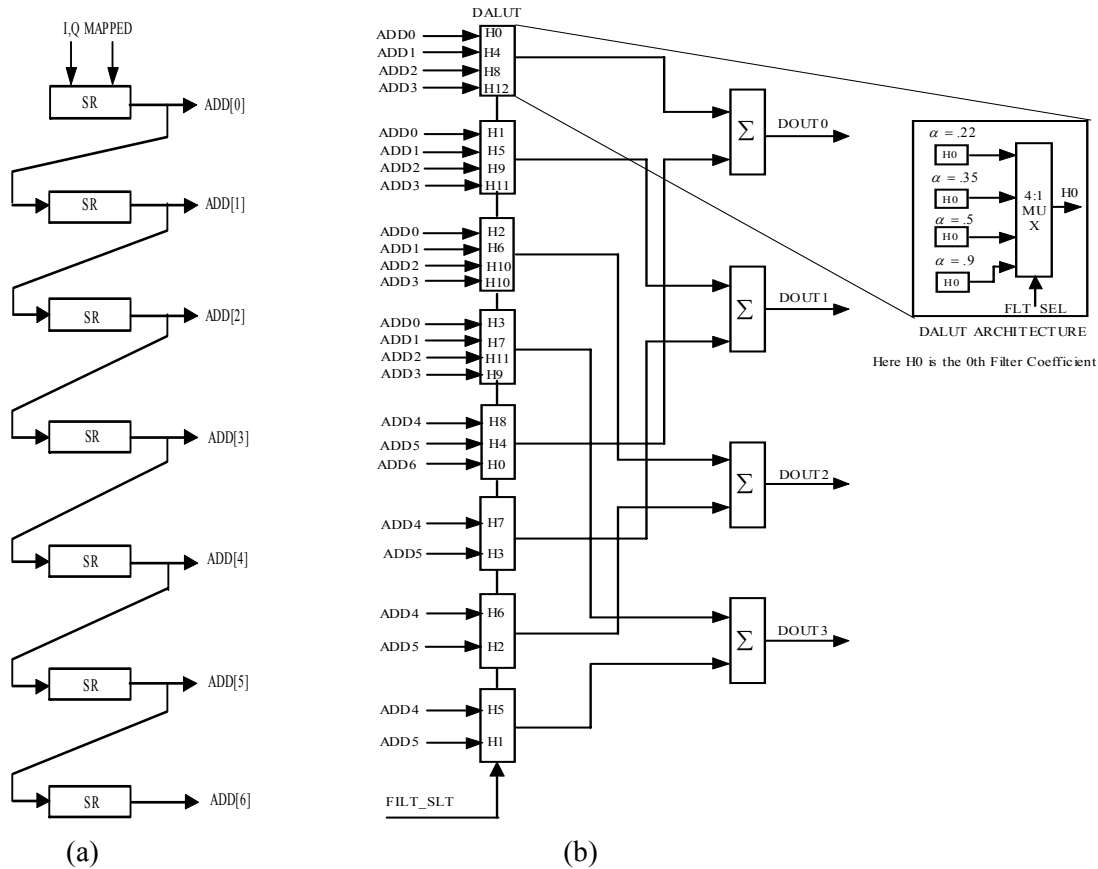

Figure 5(a) Block Diagram of Up sampler (b) Block Diagram of Designed SRRC filter

### 4.5. Carrier Generator (CG)

Complex envelope of the modulated signal can be expressed as

$$Y_K = Y_{IK} \cos w_c t + Y_{QK} \sin w_c t \tag{7}$$

where $w_c$ is the carrier frequency and I and Q denotes the in-phase and quadrature components. Equation (7) executed at every clock cycle of period (200ns). Over a symbol period of duration (800ns) only four values of the carrier occur. These can be conveniently defined as: $\cos w_c t =$ 1, 0, -1, and 0 and $\sin w_c t =$ 0, 1, 0, and -1. The modulated output does not require any explicit multiplication or addition; only a 4:1 multiplexer and a 2's complementer are sufficient. The output selects I component at the first clock cycle and Q response in the next clock cycle. This is followed by a –I response and then a –Q response, and the cycle repeat itself by choosing the IF center frequency of modulator at ¼ of the sampling rate.

## 5. FPGA IMPLEMENTATION RESULT

Area (measured in terms of hardware usage), speed (measured in terms of maximum data rate supported) and power (measured in terms of power consumed by the circuit) are some of the crucial performance metrics for any hardware implementation of SDR architecture.

### 5.1. Timing, Area, and Power Utilization Analysis Result

Xilinx ISE 9.2i [36] has been used for synthesis and implementation of the circuit. Xilinx XCV2vp30-7FF896 device has been used as the target device for FPGA implementation and XST has been used as a synthesis tool and XPower has been used for the power calculation. Simulation based power has been calculated here after applying a random data at the rate of 1.25 Mbps. The results are listed in Table 4 and Table 5 shows the timing, area and power utilization report after the FPGA implementation of the proposed modulator.

Table 4. Timing and Area Utilization Summary

| Modulation Schemes | Max. Freq. (MHz) | Area | | | |
|---|---|---|---|---|---|
| | | Slice | Slice FF | 4 input LUT | Gate count |
| QPSK with $\alpha = 0.22$ | 93.2 | 214 | 221 | 361 | 4,641 |
| DQPSK with $\alpha = 0.22$ | 166.7 | 77 | 68 | 149 | 1,489 |
| $\pi/4$ DQPSK with $\alpha = 0.35$ | 93.2 | 420 | 417 | 653 | 8,856 |
| OQPSK with $\alpha = 0.22$ | 93.4 | 219 | 225 | 367 | 4,870 |
| Proposed Modulator with $\alpha = 0.22, 0.35, 0.5, 0.9$ | 77 | 719 | 519 | 1149 | 13,032 |

It is obvious from the Table 5 that the gate count required for implementing the parameter controlled programmable modulator using the hardware multiplexing technique is substantially smaller than the linear sum of the gate count that would be required to implement the four different modulator schemes realized by fixed function technique.

Table 5. Power Utilization Summary Report

| Modulation | Total Power (mW) | Power Analysis | | | | |
|---|---|---|---|---|---|---|
| | | Clocks Power (mW) | Inputs Power (mW) | Logic Power (mW) | Ouputs Power (mW) | Signals Power (mW) |
| QPSK | 142.58 | 1.01 | 0.04 | 0.12 | 36.96 | 1.32 |
| DQPSK | 143.98 | 0.93 | 0.04 | 0.40 | 37.29 | 2.20 |
| $\pi/4$ DQPSK | 147.34 | 0.93 | 0.04 | 0.48 | 40.27 | 2.40 |
| OQPSK | 141.5 | 0.72 | 0.04 | 0.35 | 35.27 | 1.99 |
| PBM with M=0 F=0* | 140.57 | 1.16 | 0.04 | 0.11 | 34.99 | 1.14 |
| PBM with M=1 F=0* | 138.51 | 1.16 | 0.04 | 0.10 | 33.02 | 1.07 |
| PBM with M=2 F=0* | 138.83 | 1.16 | 0.04 | 0.11 | 33.32 | 1.08 |
| PBM with M=3 F=0* | 136.37 | 1.16 | 0.04 | 0.09 | 30.96 | 1.00 |
| PBM with M=0 F=1* | 140.56 | 1.16 | 0.04 | 0.11 | 34.98 | 1.14 |
| PBM with M=1 F=1* | 139.07 | 1.16 | 0.04 | 0.11 | 33.54 | 1.10 |
| PBM with M=2 F=1* | 138.66 | 1.16 | 0.04 | 0.10 | 33.16 | 1.08 |
| PBM with M=3 F=1* | 135.71 | 1.16 | 0.04 | 0.09 | 30.32 | 0.98 |
| PBM with M=0 F=2* | 140.41 | 1.16 | 0.04 | 0.11 | 34.85 | 1.13 |
| PBM with M=1 F=2* | 137.32 | 1.16 | 0.04 | 0.10 | 31.85 | 1.04 |
| PBM with M=2 F=2* | 138.74 | 1.16 | 0.04 | 0.11 | 33.23 | 1.08 |
| PBM with M=3 F=2* | 135.24 | 1.16 | 0.04 | 0.09 | 29.85 | 0.97 |
| PBM with M=0 F=3* | 138.49 | 1.16 | 0.04 | 0.10 | 33.00 | 1.07 |
| PBM with M=1 F=3* | 136.39 | 1.16 | 0.04 | 0.09 | 30.99 | 0.99 |
| PBM with M=2 F=3* | 139.25 | 1.16 | 0.04 | 0.11 | 33.72 | 1.09 |
| PBM with M=3 F=3* | 135.18 | 1.16 | 0.04 | 0.09 | 29.79 | 0.97 |

* M indicates the MOD_SEL parameter value, and F indicates the FLT_SEL parameter value.
Summing up the different power components of the proposed baseband modulator given in Table 5, the overall power comes out at maximum of 140mW.

### 5.2. Simulation Result

For the post place and route simulation in FPGA Modelsim-Xe 6.3c Starter version from Mentor graphics is used as a logic simulator. A bit stream of random signal is applied to the design and the simulation results are shown for $\pi/4$ DQPSK and DQPSK in Figure 6 and 7 respectively. In this simulation, the modulating signal at a data rate of 1.25 Mbps has been applied. The carrier frequency has been taken of 5 MHz for generating the output complex envelope. In the all the picture the signals from the top are the modulation selection signal (MOD_SEL), the RRC filter selection signal (FLT_SEL), the pulse-shaped data stream of in-phase component (IRAIL), the pulse-shaped data stream of quadrature component (QRAIL), the final modulated signal (FINALOUT).

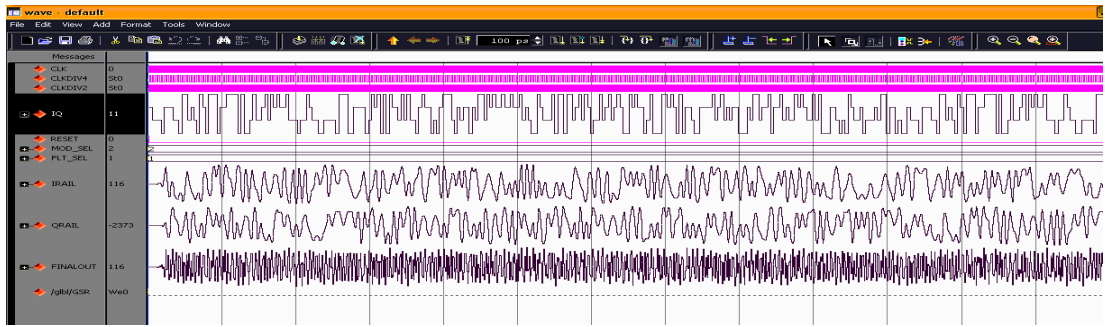

Figure 6. Simulation Result of $\pi/4$ DQPSK

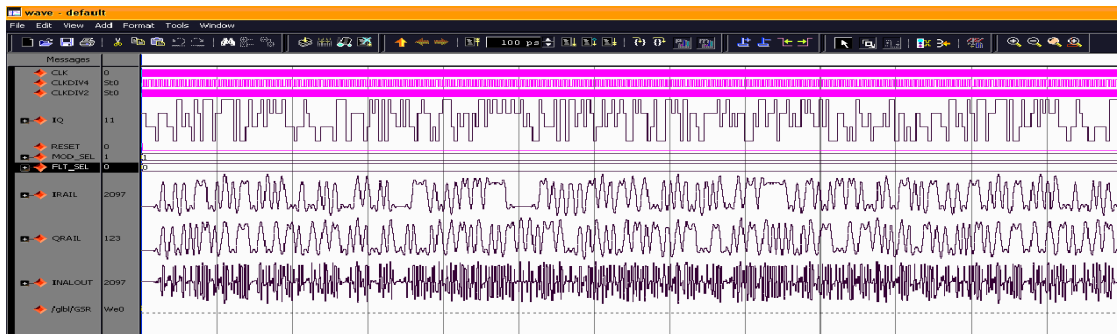

Figure 7. Simulation Result of DQPSK

### 5.3. FPGA Implementation Result

The designed system has been implemented using the Xilinx Impact tool to the Virtex-2 Pro University Board and Xilinx Chipscope-Pro 9.2i is used for capturing the modulated inphase (IRAIL) and quadrature component (QRAIL) data for verifying the FPGA implementation result of the designed circuit. Here 512 samples of the output has been captured through chipscope-pro after implementing the design into FPGA and the captured output as plotted as two-dimensional constellation diagram by setting different value of MOD_SEL are shown below in Figure 8;

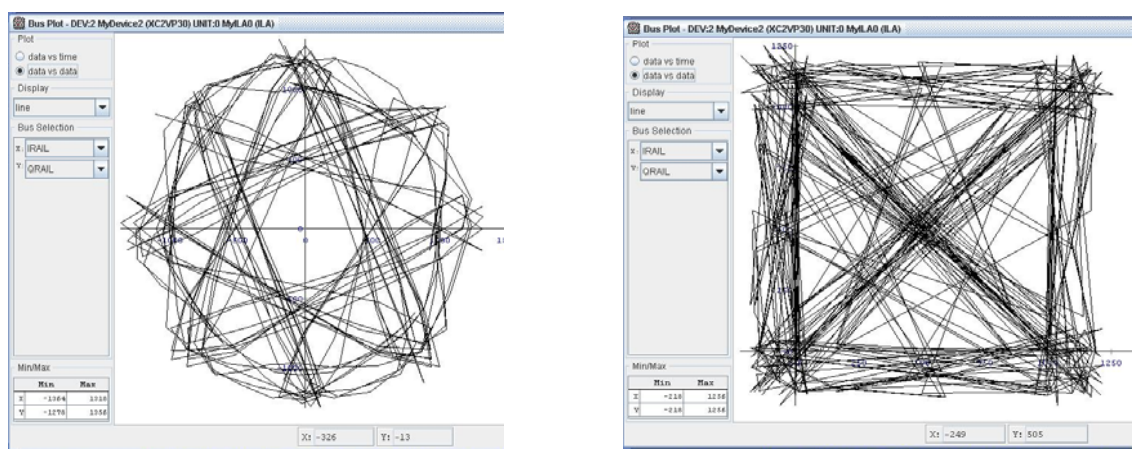

Figure 8. FPGA implementation result by setting (a) MOD_SEL=1and FLT_SEL=1 (Pi/4 DQPSK) (b) MOD_SEL=2 and FLT_SEL=1 (DQPSK)

## 6. CONCLUSIONS

Design and implementation of a parameter controlled programmable baseband modulator, which is capable of performing basic QPSK as well as its other three commonly used variant modulation schemes, is reported in the present paper. Implementation of the RRC filter which is an important constituent of the proposed baseband modulator has been carried out using the multiplier-less Distributed Arithmetic technique in order to reap the benefit of reduced power consumption. On applying a randomly generated data signal of 1.25 Mbps rate, the FPGA prototype of the baseband modulator targeted for Xilinx XCV2vp30-7FF896 device has been found to consume a moderate power of 140 mW. Hardware multiplexing technique, based on which the modulator has been designed has also contributed to low power consumption as well as lower area utilization. Based on the above observation the designed modulator may be considered appropriate for being integrated into a next generation wireless communication transmitter circuit for which the crucial objectives are low-power consumption coupled with limited area at a high data rate.

## ACKNOWLEDGEMENTS

The author would like to thank the Ministry of Communication and Information Technology, Govt. of India, New Delhi for the support and scholarship through Special Manpower Development Project (SMDP-II).

**Information of Authors**

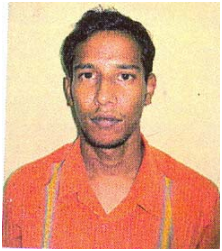

**Indranil Hatai was born in 1982, in Midnapore, India. He is currently pursuing is MS degree in Indian Institute of Technology, Kharagpur, India. His research interests include VLSI Signal Processing Architecture, Wireless Communication, and Software Defined Radio Architecture.**

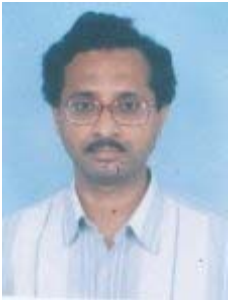

**Indrajit Chakrabarti is an Ascociate Professor in the Dept. of E&ECE of Indian Institute of Technology, Kharagpur, India. His research interest is on VLSI design and Wireless Communication.**